\documentclass[11pt,a4paper]{article}

\pdfoutput=1
\usepackage{jcappub}
\usepackage{graphicx}
\usepackage{amssymb}
\usepackage{amsmath}
\usepackage{mathtools}
\usepackage{bm}
\usepackage{latexsym}
\usepackage{epsfig}
\usepackage{float}
\usepackage{textcomp}
\usepackage{color}
\usepackage{float}
\usepackage{blindtext}
\usepackage{enumitem}
\usepackage{xcolor}
\usepackage[section]{placeins}
\usepackage{caption}
\usepackage{float}
\usepackage{siunitx}
\usepackage{tikz} 
\usepackage{pgfplots}
\usepackage{subcaption}
\usepackage[labelformat=parens,labelsep=quad,skip=3pt]{caption}
\usepackage{graphicx}
\pgfplotsset{compat=newest} 
\usepgfplotslibrary{units} 

\usepackage{calligra}
\DeclareMathAlphabet{\mathcalligra}{T1}{calligra}{m}{n}
\DeclareFontShape{T1}{calligra}{m}{n}{<->s*[2.2]callig15}{}

\usepackage[mathscr]{euscript}
\usepackage[export]{adjustbox}

\makeatletter
\gdef\@fpheader{}
\makeatother

\def\l{\left}
\def\r{\right}

\def\be{\begin{equation}}
\def\ee{\end{equation}} 
\def\bea{\begin{eqnarray}}
\def\eea{\end{eqnarray}}

\def\lsim{\mathrel{\rlap{\lower4pt\hbox{\hskip0.5pt$\sim$}}
 \raise1pt\hbox{$<$}}}         
\def\gsim{\mathrel{\rlap{\lower4pt\hbox{\hskip0.5pt$\sim$}}
 \raise1pt\hbox{$>$}}}         

\newcommand{\dd}{\mathrm{d}}

\newcommand{\Ph}{\mathcal P_h}

\definecolor{lime}{HTML}{A6CE39}
\DeclareRobustCommand{\orcidicon}{
	\begin{tikzpicture}
	\draw[lime, fill=lime] (0,0) 
	circle [radius=0.2] 
	node[white] {{\fontfamily{qag}\selectfont \tiny ID}};
	\draw[white, fill=white] (-0.0625,0.095) 
	circle [radius=0.007];
	\end{tikzpicture}
	\hspace{-2mm}
}

\foreach \x in {A, ..., Z}{\expandafter\xdef\csname orcid\x\endcsname{\noexpand\href{https://orcid.org/\csname orcidauthor\x\endcsname}
			{\noexpand\orcidicon}}
}

\title{Doubly peaked induced stochastic gravitational wave background: \\ \it{Testing baryogenesis from primordial black holes}}
\author[\ast]{Nilanjandev Bhaumik\orcidA{}
}
\author[\dagger]{, Anish Ghoshal\orcidB{}
}
\author[\dagger]{and \\ Marek Lewicki\orcidC{}
}
\affiliation[\ast]{Department of Physics, Indian Institute of Science, Bangalore 560012, India}
\affiliation[\dagger]{Institute of Theoretical Physics, Faculty of Physics, University of Warsaw,\\ ul. Pasteura 5, 02-093 Warsaw, Poland}
\emailAdd{nilanjandev@iisc.ac.in}
\emailAdd{anish.ghoshal@fuw.edu.pl}
\emailAdd{marek.lewicki@fuw.edu.pl}
\abstract{
Hawking evaporation of primordial black holes (PBHs) can facilitate the generation of matter-antimatter asymmetry. We focus on ultra-low mass PBHs that briefly dominate the universe and evaporate before the big bang nucleosynthesis. We propose a novel test of this scenario by detecting its characteristic doubly peaked gravitational wave (GW) spectrum in future GW observatories. Here the first order adiabatic perturbation from inflation and from the isocurvature perturbations due to PBH distribution, source tensor perturbations in second-order and lead to two peaks in the induced GW background. These resonant peaks are generated at the beginning of standard radiation domination in the presence of a prior PBH-dominated era. This unique GW spectral shape would provide a smoking gun signal of non-thermal baryogenesis from evaporating PBHs, which is otherwise impossible to test in laboratory experiments due to the very high energy scales involved or the feeble interaction of the dark sector with the visible sector.}

\keywords{Primordial black hole, Baryogenesis, Stochastic gravitational wave background}
\arxivnumber{}
\begin{document}
\maketitle

\tableofcontents

\section{Introduction}
\label{sec:intro}
Understanding the origin of baryon (or the matter-antimatter) asymmetry is one of the oldest open problems in physical cosmology and modern particle physics. While studies of antiparticles in cosmic rays~\cite{Cohen:1997ac}, the big-bang nucleosynthesis (BBN)~\cite{Burles:2000ju} and very precise measurements of the cosmic microwave background (CMB) radiation from Planck~\cite{Planck:2018vyg} provide a stringent constraint on the number density of baryons per entropy density, $Y_B = \frac{n_b-n_{\bar{{b}}}}{s} = 8.8\times 10^{-11}$, there is no clear evidence to indicate the origin of this asymmetry. The conventional approach to understanding baryogenesis in the early universe is based on the three well known (and necessary) Sakharov’s conditions \cite{Sakharov:1967dj}: (i) 
baryon number B-violation, (ii) violation of C- and CP-symmetries,
and (iii) particle interactions out of thermal equilibrium.

Starting from symmetric initial conditions, baryogenesis mechanisms can be classified into two broad categories, thermal and non-thermal. As the name suggests, thermal baryogenesis involves significant interactions between the dark sector and the Standard Model (SM) sector. In contrast, for non-thermal mechanisms, these interactions are either very small or zero. Non-thermal baryogenesis mechanisms are difficult to test in laboratory experiments due to their non-interactive nature. Earlier works by Hawking~\cite{Hawking:1974rv}, Carr~\cite{Carr:1976zz}, Zeldovich~\cite{Zeldovich} and others have shown that the population of ultra-low mass primordial black holes (PBH) in the very early universe can provide a viable non-thermal baryogenesis mechanism through their Hawking radiation. This scenario has been studied both in the context of grand unified theories (GUTs)~\cite{Toussaint:1978br, Barrow:1990he, Bugaev:2001xr,Baumann:2007yr,Hooper:2020otu} and for baryogenesis via leptogenesis models~\cite{Kuzmin:1985mm, FUKUGITA198645, Harvey:1990qw, Datta:2020bht, Barman:2022gjo, JyotiDas:2021shi, Barman:2021ost}. In GUT baryogenesis models, near the end of the PBH evaporation process, they emit massive particles when their horizon temperature exceeds the GUT scale. These massive particles can then further decay through out-of-equilibrium baryon-number violating processes to create the baryon-anti-baryon asymmetry in our universe, along with CP violation satisfying the Sakharov conditions~\cite{Sakharov:1967dj}. Alternatively, PBHs can emit heavy neutrinos that generate a net lepton number $L$ at lower energies. These heavy neutrinos then decay, and the lepton asymmetry gets converted to the baryon asymmetry through $B-L$ conserving sphaleron transitions.

While the non-thermal baryogenesis process leaves no direct observational imprints, the existence of PBHs and their Hawking evaporation can lead to detectable signatures in the stochastic gravitational wave backgrounds. In last few years, PBHs have gained a lot of interest due to their diverse implications on various aspects of cosmology. After the detection of very massive binary black hole mergers in LIGO-Virgo observatories~\cite{LIGOScientific:2016aoc,LIGOScientific:2020iuh, Abbott:2016blz,Abbott:2016nmj,Abbott:2017vtc}, PBHs have come up as a front-runner candidate for cold dark matter. PBHs leave imprints on various astrophysical phenomena like, gravitational waves (e.g.~\cite{Nakamura:1997sm,Clesse:2015wea,Bird:2016dcv,Raidal:2017mfl,Eroshenko:2016hmn,Sasaki:2016jop,Clesse:2016ajp,Takhistov:2017bpt}), formation of supermassive black holes~\cite{Bean:2002kx,Kawasaki:2012kn,Clesse:2015wea}, $r$-process nucleosynthesis~\cite{Fuller:2017uyd} etc. Existence of PBHs in different mass ranges is constrained considerably from LIGO-Virgo observations~\cite{Hutsi:2020sol}, the observation diffuse supernova neutrino background and gamma ray background~\cite{Carr:2020gox,  Carr:2016hva,Barnacka:2012bm, Laha:2020ivk,Laha:2019ssq, Dasgupta:2019cae, Ray:2021mxu}, gravitational microlensing observations~\cite{Niikura:2017zjd,EROS-2:2006ryy,Niikura:2019kqi}, CMB observations~\cite{Ricotti:2007au,Aloni:2016kuh,Poulin:2017bwe}, the detection of 21cm lines~\cite{Saha:2021pqf,Mittal:2021egv,Kohri:2022wzp,Hasinger:2020ptw,Tashiro:2012qe,Hektor:2018qqw}, gas heating in the interstellar medium \cite{Kim:2020ngi,Laha:2020vhg} and etc. But there is still an open window of asteroid mass range ($10^{-16}-10^{-14} M_{\odot}$) where PBHs can provide entirety of dark matter~\cite{Montero-Camacho:2019jte}.

PBHs can also form with large abundance in very low mass range ($<10^9 $ g). These ultra-low mass PBHs evaporate before BBN and therefore cannot contribute to the dark matter of our universe, but they can still dominate the universe for a short time before BBN~\cite{Allahverdi:2020bys}. In this kind of scenario, PBHs that form in the radiation-dominated epoch after inflation dominates the universe at some point, causing an early matter-dominated (eMD) era. Then they evaporate, and the standard radiation domination (RD) starts. PBHs with a very narrow mass distribution shall evaporate simultaneously, and in this case, we can take the transition from eMD-RD as a nearly instantaneous process.

In the early radiation domination (eRD) PBHs can form through various mechanisms: due to the amplified scalar curvature perturbation from ultra slow roll models of inflation~\cite{Bhaumik:2019tvl, Garcia-Bellido:2017mdw, Hertzberg:2017dkh, Ballesteros:2017fsr,Ragavendra:2020sop,Mishra:2019pzq}, warm inflation \cite{Arya:2019wck}, the first-order phase transitions~\cite{Hawking:1982ga,Kodama:1982sf,Jedamzik:1999am,Lewicki:2019gmv,Crawford:1982yz,Moss:1994pi,Freivogel:2007fx,Johnson:2011wt, Kusenko:2020pcg}, the collapse of topological defects~\cite{Hawking:1987bn,Polnarev:1988dh,MacGibbon:1997pu,Rubin:2000dq,Rubin:2001yw,Ashoorioon:2020hln,Brandenberger:2021zvn}, due to the dynamics of scalar condensates~\cite{Cotner:2016cvr,Cotner:2019ykd}, resonant reheating~\cite{Suyama:2004mz}, tachyonic preheating~\cite{Suyama:2006sr, Bassett:2000ha} etc. Large amplitude of scalar perturbations required for PBH formation, also amplify the tensor perturbation at second order and lead to detectable induced stochastic gravitational wave background (ISGWB)\cite{Saito:2008jc,Domenech:2021ztg,   Kohri:2018awv}.The effects of different reheating histories for the ISGWB have also been studied extensively \cite{Bhaumik:2020dor, Domenech:2019quo, Domenech:2020kqm}. An early matter dominated reheating phase with an instantaneous transition to radiation domination has been found to give rise to a resonant amplification in ISGWB near the cutoff scale of first order inflationary scalar power spectrum \cite{Inomata:2019ivs, Inomata:2019zqy}. The matter dominated era caused by ultra-low mass PBHs is an example of this scenario for a monochromatic mass distribution of PBHs \cite{Inomata:2020lmk}.

On the other hand, PBHs act as non-relativistic cold dark matter components. Thus for all these formation mechanisms, part of the radiation fluid is converted to non-relativistic dust matter during the PBH formation. Though the total energy density for these two fluid components stays homogeneous, the individual energy densities for both matter and radiation fluid becomes inhomogeneous. This leads to an isocurvature perturbation~\cite{Papanikolaou:2020qtd}. During PBH domination, this isocurvature perturbation contributes to the adiabatic perturbation. This first-order adiabatic scalar perturbations source the tensor perturbation at the second-order, and the ISGWB is amplified resonantly due to the sudden eMD-RD transition~\cite{Domenech:2020ssp, Domenech:2021wkk}.

These two contributions of induced gravitational waves have been treated separately in previous works. Though this treatment is justified for first-order scalar perturbation calculation, this assumption can be erroneous for calculating second-order tensor perturbation, mainly when these contributions fall nearly in the same frequency range. In this work, we treat these two components in a unified way and explore the corresponding ISGWB spectra for the possible detection in future GW observatories. The characteristic two-peaked shape of the ISGWB spectrum is a unique signature of the baryogenesis scenario induced by the PBH-evaporation. Confirmation of these baryogenesis processes in laboratory experiments is otherwise difficult.

The rest of the paper is organized as follows: in section~\ref{BB} we calculate the dynamics of various background quantities in the three-phase model. In section~\ref{baryogenesis1}, we briefly revisit the details of baryogenesis calculation and in section~\ref{GWPBH} we discuss the details of our numerical setup for estimating resonant ISGWB, for both isocurvature induced adiabatic and inflationary adiabatic perturbations. Then we present approximate analytical formulae to explore the possibility of ISGWB detection in section~\ref{Anum} and derive combined constraints on PBH mass range and initial abundance ($\beta_f$) in section~\ref{combined}. Finally, we summarize our findings and discuss further possibilities in section~\ref{discuss}.



\section{Background dynamics with a sandwiched PBH dominated era}
\label{BB}

For a universe dominated by fluid with a general equation of state parameter $w$,  it is possible to express the scale factor using the Friedmann equations and continuity equation\footnote{We use $c=\hbar=M_{Pl}=1$, and the mostly positive sign convention (-,+,+,+) for the metric. Derivative with respect to physical time($t$) is denoted with an overdot and derivative with respect to conformal time( $\tau$) with prime.}
\begin{align}
    a(\tau)=\left( c_0 \tau + c_1\right)^{\frac{2}{3 w+1}}, \hspace{2cm}  a(t)=\left( c_2 t + c_3\right)^{\frac{2}{3 (w+1)}} \, .
\end{align}
Lets assume PBHs are formed during early radiation dominated era at $\tau=\tau_f$, which then dominates the universe at $\tau=\tau_{m}$ and evaporates at $\tau=\tau_{r}$. We have three distinct epochs to think about: early radiation domination(eRD), PBH dominated or early matter domination (eMD) and then the standard radiation domination (RD). We get the Hubble parameter and the scale factor as a function of conformal time $\tau$ in these three phases,
\begin{align}
\label{scaleH}
    a_{eRD}(\tau)=\left(\frac{a_f}{\tau_f}\right)\tau \text{,} \hspace{0.5cm}  a_{eMD}(\tau)=\frac{a_f(\tau + \tau_{m})^2}{4\tau_f\tau_{m}}\text{,}\hspace{0.5cm} a_{RD}(\tau)=\frac{a_f(\tau_{r} + \tau_{m})(2\tau-\tau_{r} + \tau_{m})}{4\tau_f\tau_{m}} \\
    H_{eRD}(\tau)=\frac{\tau_f}{a_f \tau^2}\text{,} \hspace{0.5cm}  H_{eMD}(\tau)=\frac{8 \tau_f \tau_{m}}{a_f (\tau + \tau_{m})^3}, 
    \hspace{0.5cm}  H_{RD}(\tau)=\frac{8 \tau_f \tau_{m}}{a_f (2 \tau + \tau_{m}-\tau_{r})^2(\tau_{r} + \tau_{m})}. \label{scaleaH}
\end{align}
As the conformal time increases monotonically throughout all these three phases, expecting $\tau_f\ll\tau_{m}\ll\tau_{r}$, we get $H_{eRD} \propto \tau^{-2}$, $H_{eMD} \propto(\tau + \tau_{m})^{-3} $ and $ H_{RD} \propto 1/ (2 \tau - \tau_{r})^2$. We can then track the evolution of the Hubble parameter to infer the energy density of the universe at the time of PBH formation,
\begin{align}
\rho_{f}=\frac{\rho_{EQ}\tau_{EQ}^4\tau_{r}^2}{4 \tau_f^4\tau_{m}^2}=3 H_f^2 M_{\rm Pl}^2 ~,
\label{rhof}
\end{align}
where we assume the energy density of the universe $\rho=\rho_{EQ}$ at $\tau=\tau_{EQ}$ when the standard RD era ends and late matter domination starts. We can also express it in terms of background variables using equation \eqref{scaleH},
\begin{align}
\Delta t_{\rm PBH}=\int_{t_f}^{t_{r}} dt=\int_{\tau_f}^{\tau_{r}} a(\tau) d\tau=\int_{\tau_f}^{\tau_{m}} a_{eRD}(\tau) d\tau+\int_{\tau_{m}}^{\tau_{r}} a_{eMD}(\tau) d\tau \approx \frac{a_f \tau_{r}^3}{12 \tau_f  \tau_{m}} ~.
\label{confcon}
\end{align}
 As the mass fraction of PBH varies proportional to the scale factor during eRD, it is possible to express the initial mass fraction of PBHs, $\beta_f=\tau_f/\tau_{m}$. We assume instantaneous collapse of large over-densities to form PBHs just after the horizon entry of relevant perturbation modes. Therefore the initial PBH mass, $M_{\rm PBH}$ and Hubble horizon can be connected as, $M_{\rm PBH}=\gamma 4\pi M_{Pl}^2/H_f$. Here $\gamma \approx 0.2$ is an efficiency factor for PBH collapse in eRD. Solving equation~\eqref{rhof} and~\eqref{confcon} we get,
 \begin{align}
 \label{tauRD}
\tau_{r}=\sqrt{2} \left(\frac{3\Delta t_{\rm PBH}^2\rho_{EQ}\tau_{EQ}^4}{M_{Pl}^2}\right)^{1/4} \, .
\end{align}
From equation \eqref{rhof}, we can also calculate the ratio of two conformal times, $\tau_r$ and $\tau_m$,
 \begin{align}
\tau_{rat} \equiv \frac{\tau_{r}}{\tau_{m}}= &2\left(\frac{3\pi^2 \gamma^2 M_{Pl}^6 \beta_f^4 \tau_{r}^4 }{M_{\rm PBH}^2 \rho_{EQ}\tau_{EQ}^4 } \right)^{1/6}\nonumber\\
=& 2 (6 \pi \gamma)^{1/3} \left( \frac{\beta_f^4 \Delta t_{\rm PBH}^2 M_{Pl}^4}{M_{\rm PBH}^2} \right)^{1/6} \, .
\label{taurat}
\end{align}
The relevance of this ratio lies in the validity of the linear theory. The linear approximation for the scalar perturbation, does not hold for $\tau_{rat} > 470$ ~\cite{Inomata:2019zqy, Kohri:2018awv, Assadullahi:2009nf}.

For a chargeless non-rotating Schwarzschild black hole the horizon temperature can be written as~\cite{peacock_1998, Hawking:1975vcx},
\begin{equation}
T_{\rm BH} = \frac{M_{Pl}^2}{
M_{\rm PBH}} \, ,
\end{equation}
the rate of mass change per unit time as,
\begin{equation}
\frac{d M_{\rm PBH}}{d t} = -\frac{ \pi~ \mathcal{G}~ g_{*, H} M_{Pl}^4}{480  M_{\rm PBH}^2} \, .
\end{equation}
and mass of a PBH determines the physical time interval between its formation and evaporation, $\Delta t_{\rm PBH}$,
\begin{equation}
\label{tevp}
    \Delta t_{\rm PBH}=\frac{160  M_{\rm PBH}^3}{\pi~ \mathcal{G}~ \overline{g_{*,H}} ~M_{Pl}^4 } \, .
\end{equation}
Here $\mathcal{G} \approx 3.8$ is the graybody factor,  $g_{\star, H}$  is the number count of degrees-of-freedom for particles with masses below $T_{BH}$ and $\overline{g_{*,H}}$ is average over PBH lifetime. We take $\overline{g_{*,H}} \approx 108$ considering only Standard Model particles and assuming $M_{\rm PBH} < 10^9 ~\text{g}$  \cite{Hooper:2020otu,MacGibbon:1991tj} .
For each conformal time $\tau_Y$ we get a comoving wavenumber $k_Y\equiv 1/\tau_Y$ which re-enters the horizon at $\tau=\tau_Y$. Using equation  \eqref{tauRD}, \eqref{taurat}, \eqref{tevp} and taking $k_{EQ}=1/\tau_{EQ}=0.01 \text{ M}pc^{-1}$ and $H_{EQ}=20.7 \text{ M}pc^{-1}$ we find,
\begin{align}
k_{r} & = 1/\tau_r \approx 2.1 \times 10^{11} \left( \frac{M_{\rm PBH}}{10^4 {\rm g} } \right)^{-3/2} \text{ M}pc^{-1} , \\
 k_{m} & = \tau_{rat}/\tau_r \approx 3.4 \times 10^{17} \left( \frac{M_{\rm PBH}}{10^4 {\rm g} } \right)^{-5/6} \beta_f^{2/3} \text{ M}pc^{-1},  \\
\label{k_f}
\hspace{2cm}  k_{f} & = \frac{k_{m}}{\beta_f}\approx3.4 \times 10^{17} \left( \frac{M_{\rm PBH}}{10^4 {\rm g} } \right)^{-5/6} \beta_f^{-1/3} \text{ M}pc^{-1} .
\end{align}
One interesting point to note from this section is that $k_{r}$ does not depend on $\beta_f$, despite the fact that both $k_{m}$ and $k_{f}$ explicitly depend on $\beta_f$. In our study, these three wavenumbers play a very crucial role. The wavenumber $k_r$ associated with the eMD-RD transition indicates when the ISGWB is generated, baryogenesis happens. Inflationary scalar perturbation modes with wavenumber $k > k_m$ enter during eRD and get highly suppressed by the time PBHs evaporate. We shall see in later sections that, in our case, the majority of ISGWB contribution is generated just after the PBH evaporation. Therefore, $k_m$ acts as the cutoff scale for inflationary scalar perturbations. For the power-law power spectrum, this cutoff scale leads to the first resonant peak in ISGWB. The scale of PBH formation $k_f$ is directly associated with the mass of PBHs and plays a crucial role in determining the cutoff scale for isocurvature perturbation. We will discuss this at greater length in section~\ref{GWPBH}.

\section{Baryogenesis from ultra-low mass PBHs}
\label{baryogenesis1}
 We consider two scenarios of baryogenesis. One involves direct baryogenesis through the decay of Higgs triplets. In another case, baryogenesis comes through leptogenesis due to the decay of right-handed neutrinos. Both Higgs triplet and right-handed neutrinos can come from Hawking evaporation of PBHs. The emission of these particles occurs when the PBH horizon temperature becomes higher than the energy scales or the masses of these particles. Here PBH mass range sets the scale of baryogenesis and leptogenesis, which can even be arbitrarily high up to the GUT scale. For GUT models, massive Higgs triplets are emitted from PBH and subsequently decay with baryon number violating processes to directly generate the observed matter-antimatter asymmetry, fulfilling all the Sakharov's conditions.
On the other hand, right handed neutrinos generate a net lepton number $L$ at lower energy. This lepton asymmetry gets converted to the baryon asymmetry through $B-L$ conserving sphaleron transitions.  We shall denote both Higgs triplets and right handed neutrinos with $X$ ($X= \cal T \text{~and~} \cal N$). For both these cases, we associate an efficiency factor $\epsilon_X$\cite{Hooper:2020otu},
\begin{equation}\label{epsilon}
\epsilon_X \equiv \sum_i B_i \frac{\Gamma (X\to f_i) - \Gamma(\bar{X} \to 
\bar{f}_i)}{\Gamma_{\rm tot}}\, .
\end{equation}
Here $f_i$ denotes $i^{th}$ final particle, with baryon number $B_i$ , and $\Gamma_{tot}$ is the total width for the decay of Higgs triplet. For the second scenario, $\epsilon_X$ corresponds to the efficiency of the combined leptogenesis and baryogenesis processes. 

The rate of production of a heavy particle $X$ with energy $E_X$, per unit black hole surface area can be written as,
\be
\frac{dN_{ X}}{ dt \, dA} = \frac{ {\cal G} g^{X}_{H}}{4} \int \frac{d^3p}{(2\pi)^3} \frac{1}{e^{E_X/T_{\rm BH}} \pm 1},
\ee
where $-$ and $+$ sign corresponds to bosons and fermions respectively and $g^{X}_{H}$ denotes the number of degrees of freedom associated with particle $X$. In order to estimate the total number of particles $N_{X}$ emitted by the PBH, we integrate from the initial black hole temperature $T_{{\rm BH,}i} =M _{\rm Pl}^2/M _{\rm PBH}$ to $M_{Pl}$,
\be
\label{NT}
N_{X}  =
 \frac{ 30  M^2_{\rm Pl}}{  \pi^2    }
\int^{M_{\rm Pl}}_{T_{{\rm BH,}i}} \frac{  dT_{\rm BH}  }{T^6_{\rm BH}}
\left( \frac{ g^{X}_{H}}{ g_{\star, H}} \right)
 \int \frac{d^3p}{(2\pi)^3} \frac{1}{e^{E_X/T_{\rm BH}} \pm 1}\, .
\ee
The resulting baryon asymmetry from the decay of these heavy particles can be written as,
\be
\label{BHdom}
Y_B = \frac{n_{\rm PBH}}{s} \epsilon_{X} N_{X}\, ,
\ee
where $n_{\rm PBH}$ is the number density of PBHs and $s$ is the entropy density at the time of PBH evaporation. During the evaporation, number density of black holes $n_{\rm PBH}$ can be written as, 
\be
n_{\rm PBH}=\frac{\rho_{\rm PBH}}{M_{\rm PBH}}=\frac{3 H_r^2 M_{\rm Pl}^2}{M_{\rm PBH}}\, .
\ee
The Hubble parameter after the evaporation or at the start of the RD can be written as  $H_r=2/(3 \Delta t_{\rm PBH})$. We have defined  the lifetime of PBHs $\Delta t_{\rm PBH}$ in equation \eqref{tevp}. If the decay process is very efficient, heavy decaying particles can lead to feasible baryogenesis scenario and we can stay in $T_{\rm BH, i} \ll M_{X}$ limit. Then the expression for $Y_B$ is given by~\cite{Hooper:2020otu},
\be
\label{BHdom2a}
Y_B \approx  2.1 \times 10^{-10}  \,  
 \bigg(\frac{10^{12}\,{\rm GeV}}{M_{X}}\bigg)^2 \, \bigg(\frac{10^2 \, {\rm g}}{M_{{\rm PBH}} }\bigg)^{5/2}  \bigg(\frac{\epsilon_{X}}{10^{-2}}\biggr) 
 \bigg(\frac{90}{g_{\star}}\bigg)^{1/4}\, .
\ee
We assume the number of degrees-of-freedom of relativistic particles during RD epoch $g_{*}$ to be $\approx 106.7$. Taking $Y_B  \approx 8.8 \times 10^{-11} $ from equation \eqref{BHdom2a} it is possible to obtain $M_{X}$ as a function of efficiency parameter $\epsilon_X$ and initial PBH mass $M_{\rm PBH}$,
\be
\label{MX}
M_{X} \approx 6.34 \times 10^{15} \sqrt{\epsilon_X\left(\frac{1 ~\text{g}}{M_{\rm PBH}}\right)^{5/2}} ~\text{GeV} \, .
\ee
Now in GUT model, the mass of the Higgs triplet particles $M_{X} \gsim 3 \times 10^{11} \, {\rm GeV}$, as required by the laboratory constraints on proton decay from SuperK experiment~\cite{Super-Kamiokande:2016exg}. Thus to find a bound on the PBH mass in the Higgs triplet case, one can further approximate~\cite{Barrow:1990he,Baumann:2007yr,Morrison:2018xla},
\be
\label{BHdom3}
  M_{X} \gsim 3 \times 10^{11} \text{GeV} \implies M_{\rm PBH} \lsim 5 \times 10^2 \, \left( \frac{ \epsilon_{X}}{10^{-2}  } \right)^{2/5}  \, {\rm g} \, .
 \ee
 It is important to note that, while equation~\eqref{MX} is valid for both Higgs triplet and right-handed neutrinos, the equation~\eqref{BHdom3} is only valid in the case of Higgs triplet particles.

\section{Two peaks of the induced stochastic gravitational waves background (ISGWB)}
\label{GWPBH}
In this section we will discuss the production mechanisms behind the two peaks of the GW spectrum and the method of combining them into our final result.

\subsection{Isocurvature and adiabatic perturbations from PBH dominated universe}
\label{PPS}
Ultra-low mass PBHs ($M_{\rm PBH} < 10^9$ g) evaporate due to Hawking radiation before the BBN, and as a result, the abundance of such black holes is essentially unconstrained. However, they can dominate the universe for a brief period before the standard radiation domination (RD) begins from the decay products of their Hawking evaporation. PBHs are distributed randomly, and their energy density (redshifting like matter) is inhomogeneous. During PBH formation, radiation fluid converts to non-relativistic PBH dark matter. Thus, the total energy density of PBH and radiation fluid remains homogeneous for scales larger than the PBH-forming scale. Still, the individual energy density of either radiation fluid or PBHs becomes inhomogeneous. Recently it was shown that this inhomogeneity leads to initial isocurvature perturbations~\cite{Papanikolaou:2020qtd}, which later convert to curvature perturbations and contribute to ISGWB. This ISGWB can constrain the initial abundance of ultra-low mass PBHs~\cite{Dom_nech_2021, Domenech:2020ssp}.
Assuming a Poissonian distribution of primordial black holes one can infer the initial isocurvature  power spectrum~\cite{Papanikolaou:2020qtd},
\begin{align}
    \mathcal{P}_S(k,\tau_f)= \frac{2}{3\pi} \left(\frac{k}{k_{\rm UV}}\right)^3\, .
\end{align}
We are taking the cutoff of the powerspectra at $k_{\rm UV}$, scale of mean distance between two black holes( $\overline{r}_f$) at their formation, 	
\begin{align}
    k_{\rm UV} &= \frac{a_f}{\overline{r}_f}\nonumber \\
    &=\gamma^{-1/3}\beta^{1/3}a_f H_f =\gamma^{-1/3}\beta^{1/3} k_f\, .
\end{align}
 For smaller scales ($ k> k_{\rm UV}$) we cannot ignore the effects of granularity in the PBH fluid. We derived $k_f \propto \beta_f^{-1/3} $ in  equation  \eqref{k_f}. Therefore the cutoff scale for isocurvature perturbation $k_{\rm UV}$ does not depend on initial abundance of PBHs $\beta_f$. During eRD and then during PBH dominated eMD, this isocurvature perturbation will convert to the adiabatic perturbations. Thus after the evaporation of PBHs at the beginning of the standard radiation domination ($\tau=\tau_r$), we can write the Bardeen potential, $\Phi(k)$ as a sum of the two components,
\begin{align}
 \Phi(k,\tau_r)=\Phi_{\rm infl}(k,\tau_r)+\Phi_{\rm PBH}(k,\tau_r)\, .
\end{align}
While the first component comes from inflationary scalar perturbations, the second component comes due to the isocurvature perturbations introduced by PBHs. At linear order it is justified to trace the evolution of these two components separately. Assuming the initial value of the $\Phi_{\rm PBH}(k)$ at the PBH formation epoch to be zero we further trace the  evolution of adiabatic perturbations in presence of isocurvature perturbations. Neglecting the effects of peculiar velocities of PBHs  we can calculate the power spectrum of the second component \cite{Papanikolaou:2020qtd},
\begin{align}
    \mathcal{P}_{\Phi_{\rm PBH}}(k,\tau_r)= \frac{2}{3\pi} \left(\frac{k}{k_{\rm UV}}\right)^3  \left(5+\frac{4}{9}\frac{k^2}{k_{m}^2}\right)^{-2}\, .
\end{align}
Here $k_{m}$ denotes the wavenumber of the mode which re-enters the horizon at the onset of transition from early-radiation to PBH dominated era. It is important to note that, we are calculating the power spectrum at the beginning of second stage of RD. As PBH evaporation is not a perfectly instantaneous process, the finite duration will affect the modes whose time variation is larger than the rate of transition. This gives rise to an additional  k-dependent suppression factor, largely affecting the isocurvature induced part~\cite{Domenech:2020ssp}. If we assume that the PBH evaporation starts at $\tau=\tau_{r0}$ and completes at  $\tau=\tau_{r}$, we have,
\begin{align}
  \frac{\Phi(k,\tau_{r})}{\Phi(k,\tau_{r0})} \approx \left(\sqrt{\frac{2}{3}} \frac{k}{k_r} \right)^{-1/3}\, .
\end{align}

Earlier works in this direction ignore the contribution from already existing adiabatic perturbations, assuming that it is possible to treat these two contributions separately in linear theory. 
For the estimation of ISGWB we extend their formalism taking into account both inflationary and isocurvature induced adiabatic perturbations from the distribution of PBHs. Assuming that the isocurvature-induced adiabatic and inflationary adiabatic components are completely uncorrelated, we can neglect the cross term and the scalar power spectrum for the combined contribution can be written as,
\begin{align}
    \mathcal{P}_{\Phi}(k,\tau_r)= \mathcal{P}_{\Phi_{\rm PBH}}(k,\tau_r)+ \mathcal{P}_{\Phi_{\rm infl}}(k,\tau_r)
    \, .
\end{align}
We consider standard power-law power spectra for inflationary scalar curvature perturbations, 
\begin{align}
\label{power-law}
\mathcal{P}_{\mathcal{R}}=A_s \left( \frac{k}{k_{p}} \right)^{n_s -1}.
\end{align}
with pivot scale $k_p=0.05 \, {\rm Mpc}^{-1}$, scalar amplitude $A_s= 2.09 \times 10^{-9} $ and scalar index $n_s=0.965$ \cite{Planck:2018jri}. As we are calculating the power spectrum at the beginning of second stage of RD, we cannot directly use inflationary power spectrum of comoving curvature perturbation, $\mathcal{P}_{\cal R}(k)$ to get $\mathcal{P}_{\Phi_{\rm infl}}(k,\tau_r)$. The inflationary scalar perturbation modes ( $k > k_m$) that re-enters the horizon during eRD, gets significantly suppressed due to their evolution during eRD. This sets the cutoff scale of $\Phi_{\rm infl}(k,\tau_r) $ at $k=k_m$. 


\subsection{ ISGWB from the two contributions}
\label{secondarygw}

\begin{figure}[h!]
\begin{center}
\includegraphics[width=15.5cm]{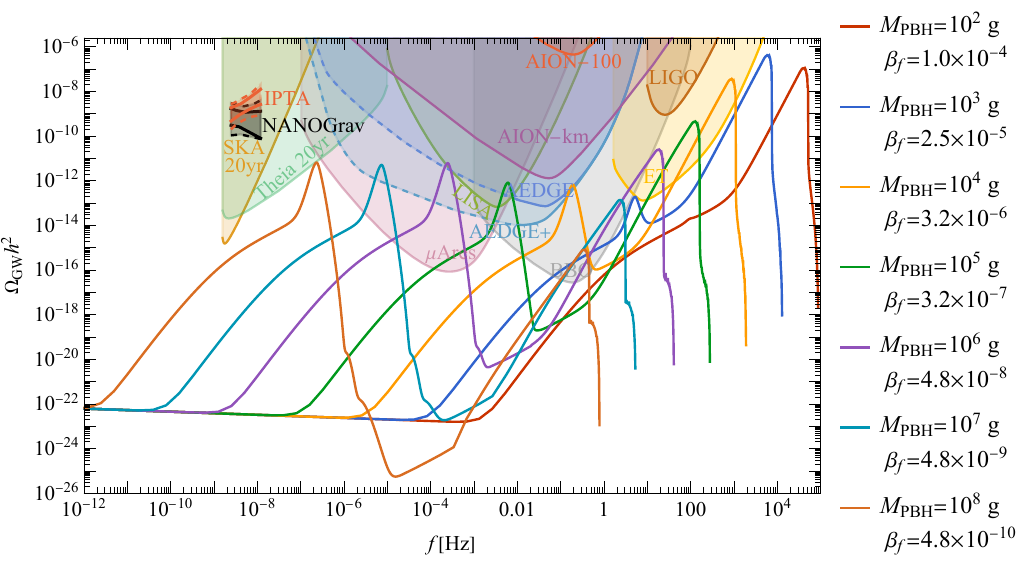}
\caption{\it Spectral energy density of Induced stochastic GW background for a range of primordial black hole populations. In each case the PBHs dominate the expansion of the universe for a time before they evaporate. $\beta_f$ denotes the initial mass fraction of PBHs at their formation.
The two peak profiles correspond to the secondary contribution sourced by inflationary adiabatic and isocurvature induced adiabatic scalar perturbations. As we have discussed, the adiabatic contribution peaks around $k_{m}$ while the isocurvature contribution peaks at $k_{\rm UV}$ the cutoff scale of the isocurvature power spectrum.}

\label{sps-gw1}
\end{center}
\end{figure}

Though at the linear order  the scalar, vector and tensor perturbations evolve independently, these three components cannot be decoupled so easily at second order. The amplification of second order tensor perturbations sourced by the amplified first order scalar perturbations has been studied extensively in the context of primordial black hole formation ~\cite{Kohri:2018awv, Bartolo:2018rku, Espinosa:2018eve}. In order to estimate the induced stochastic gravitational wave background at second order we closely follow the formulation introduced in~\cite{Bhaumik:2020dor}. In the conformal Newtonian gauge, we can write~\cite{Bardeen:1980kt}
\begin{equation}
ds^2 = -a^2(\tau)(1+2\Phi)\, d\tau^2 + a^2(\tau) \left[(1-2\Psi)\delta_{ij}+\frac{1}{2}h_{ij} \right] dx^i dx^j\, ,
\end{equation}
where $\Phi$ and $\Psi$ are the scalar metric perturbations, or the Bardeen potentials and $h_{ij}$ is the traceless and transverse tensor perturbation components. Using the standard  quantisation procedure for the Fourier modes $h_{\bf k}$ we get,
\begin{equation}
h_{\bf k}''(\tau) + 2 \mathcal{H} h'_{\bf k}(\tau)+ k^2 h_{\bf k}(\tau) = 4 S_{\bf k}(\tau)\, ,  \label{EOM_h}
\end{equation}
where $S_{\bf k}$ is the Fourier component of the source term coming from first order scalar perturbations. We can solve this equation using Green's function method which allows us to write the power spectrum of tensor perturbations~\cite{Matarrese:1997ay, Ananda:2006af, Baumann:2007zm},
\begin{equation}
\Ph(\tau,k) = 4 \int_{0}^{\infty}\dd v \int_{|1-v|}^{1+v} \dd u
\l(\frac{4 v^2-(1+v^2-u^2)^2}{4 u v}\r)^2 I^2(v,u,x) {\cal P}_{\cal R}(kv){\cal P}_{\cal R}(ku) \, .
\label{P-h}
\end{equation}
In this convolution integral, the last two terms represent the initial power spectrum of first-order scalar perturbations and the factor $I^2(v,u,x)$ takes into account the time evolution of scalar perturbations. In our case, in the context of an intermediate phase of PBH domination, we can break the contributions of $I(v,u,x)$ into 3 different components, 
\begin{align}
    I(v,u,x)=I_{\rm eRD}+I_{\rm PBH}+I_{\rm RD}\, .
\end{align}
These components correspond to the generation of ISGWB in three different phases: first, the eRD phase, followed by the PBH-dominated eMD phase, and finally, the late RD. For the isocurvature induced gravitational wave component, it has been shown that the dominant contribution to the gravitational wave spectrum comes due to the resonant contribution~\cite{Inomata:2019ivs} at the very beginning of standard RD~\cite{Domenech:2020ssp}. For standard power-law power spectra the inflationary scalar curvature perturbations also contribute dominantly only at the start of RD phase\cite{Inomata:2020lmk}. Thus, our calculations only focus on the gravitational waves generated during the standard RD phase \cite{Inomata:2020lmk}. We use a dimensionless variable $x$, a product of conformal time $\tau$ and comoving wavenumber $k$. Here $x_r=k\tau_r$ corresponds to the beginning of the RD phase, and $x=k\tau$ corresponds to some late time $\tau$ during the RD epoch by when the source function would stop contributing. This allows us to express,
\begin{equation}{\label{IIR}}
\begin{aligned}
I(v,u,x) \approx I_{\rm RD}(u,v,x)= & \int_{x_r}^x \dd\bar{x} ~\frac{a_{\rm RD}(\bar{x})}{a_{\rm RD}(x)}~f(u,v,\bar{x},x_r) ~k~ G(\bar{x},x)\, ,
   \end{aligned}
\end{equation}
where
\begin{align}
\label{RD-Omega}
\frac{a_{\rm RD}(\bar{x})}{a_{\rm RD}(x)}=&\frac{\bar{x}-x_r/2}{x-x_r/2} \, ,
\end{align}
and assuming the peculiar velocities of
PBHs to have negligible contribution we can calculate $f(u,v,\bar{x},x_r)$ in terms of transfer function $\mathcal{T}$ and its time derivative \cite{Bhaumik:2020dor}, 
\begin{align}\label{ff}
f(u,v,\bar{x},x_r)=\frac{4}{9} \biggl[ (\bar{x}-x_r/2) \partial_{\bar{x}}\mathcal{T}(u \bar{x},u x_r) ( (\bar{x}-x_r/2)\partial_{\bar{x}}\mathcal{T}(v \bar{x},vx_r) +\mathcal{T}(v \bar{x},v x_r)) \biggr. \nonumber \\
+ \biggl. \mathcal{T}(u \bar{x},u x_r) ((\bar{x}-x_r/2)
\partial_{\bar{x}}\mathcal{T}(v \bar{x},v x_r)+3 \mathcal{T}(v \bar{x},v x_r)) \biggr].
\end{align}
As we limit ourselves to a finite duration of PBH dominated era we can assume $\tau_{m} \ll \tau_{RD}$, and from equation~\eqref{scaleaH} we can take the scale factor during RD $a_{RD} \propto (\tau - \tau_r/2)$. In RD, ${\cal H}=a H=1/(\tau - \tau_r/2)$, and we can write $k/{\cal H}=k(\tau - \tau_r/2)=(x-x_r/2) $. Now we can define,
 \begin{equation}{\label{CI}}
\mathcal{I}(u,v,x,x_r)=I(u,v,x,x_r)\times (x-x_r/2) \, ,
\end{equation}
and express $\Omega_{GW}(\tau,k)$ as,  
\begin{align}
\Omega_{GW}(\tau,k) &= \frac{1}{24} \left( \frac{k}{{\cal H}} \right)^2 \overline{\mathcal{P}_h(\tau, k)}\\
&= \frac{1}{6}  \int_{0}^{\infty}\dd v \int_{|1-v|}^{1+v} \dd u
\l(\frac{4 v^2-(1+v^2-u^2)^2}{4 u v}\r)^2  \overline {\cal I}_{\rm RD}^2(v,u,x) {\cal P}_{\cal R}(kv){\cal P}_{\cal R}(ku)\, . 
\label{omega-in-u-v}
\end{align}
 Due to the presence of an intermediate PBH dominated phase, we cannot use the standard pure RD expression for  $\overline {\cal I}_{\rm RD}^2$. Instead, we use the exact full expression for  $\overline {\cal I}_{\rm RD}^2$  derived in the appendix A of ref.~\cite{Bhaumik:2020dor} for our numerical code. We are assuming the peculiar velocities of PBHs to have negligible contribution in $\overline {\cal I}_{\rm RD}^2$ compared to other terms. 
 
 After the source function stops to contribute, the fraction of energy density associated with ISGWB stays constant until the RD phase ends at $\tau=\tau_{EQ}$. Then during late MD and dark energy domination, $\Omega_{GW}$ scales like the radiation component. Using the entropy conservation, we can express the present ISGWB energy density as~\cite{Espinosa:2018eve} 
\bea
\Omega_{GW}(\tau_0,k) =  c_g ~\Omega_{r,0} ~\Omega_{GW}(\tau,k)\, ,
\eea
where $\Omega_{r,0}$ is the present radiation energy density, $c_g \approx 0.4$ if we take the number of relativistic degrees of freedom to be $\sim 106.7$. 

Fig.~\ref{sps-gw1} shows the GW spectra for PBH mass range $10^2 ~-~10^8$ g with appropriate mass fractions.
For reference we also show the projected Power-Law integrated sensitivities~\cite{Thrane:2013oya} of LIGO~\cite{LIGOScientific:2014pky,LIGOScientific:2016fpe},
 SKA~\cite{Janssen:2014dka}, LISA~\cite{Bartolo:2016ami,Auclair:2022lcg},  AEDGE~\cite{Badurina:2021rgt,AEDGE:2019nxb}, AION/MAGIS~\cite{Badurina:2021rgt,Badurina:2019hst,Graham:2016plp,Graham:2017pmn}, ET~\cite{Punturo:2010zz,Hild:2010id}, BBO~\cite{Yagi:2011wg,Crowder:2005nr},
 $\mu$ARES\cite{Sesana:2019vho},
 and Theia\cite{Garcia-Bellido:2021zgu}. 
 We also show fits to the hints for a possible signal reported by PTA collaborations~\cite{NANOGrav:2020bcs,Goncharov:2021oub,Chen:2021rqp,Antoniadis:2022pcn}.
The impact of astrophysical foregrounds, for example, from the population of BH currently probed by LIGO and Virgo on the reach of the experiments is likely to be important~\cite{Lewicki:2021kmu}. However, a dedicated analysis would be required to precisely measure the impact on our spectra, and we leave this problem for further studies. 

The origin of the amplification in this scenario is the resonant contribution generated at the beginning of RD, coming solely due to the non-trivial evolution of scalar perturbation modes during the previous PBH-dominated eMD era, as elaborated in earlier work by Inomata et al. \cite{Inomata:2019ivs}. The perturbation modes that re-enter during eRD or eMD go through a matter-dominated phase when the first-order scalar perturbation stays nearly constant and starts to oscillate and decay at the very onset of RD. This oscillation contributes dominantly to the product of temporal derivative term in \eqref{ff}. The amplitude of this contribution depends both on the amplitude of the scalar mode and the wavenumber of the mode. The perturbation modes that re-enter during eMD or eRD correspond to higher wavenumbers than the conformal Hubble parameter at the start of RD. Higher wavenumber modes oscillate with higher frequency and contribute more. Thus, we obtain the peaks in ISGWB corresponding to the scalar power spectrum cutoff scales.

As we have discussed in section \ref{BB}, for an intermediate PBH domination, the effective cutoff for the inflationary power spectrum corresponds to the mode, $k_m$, which re-enters at the start of the eMD. Any mode with a higher wavenumber ($k > k_m$) enters during eRD and gets suppressed during their evolution in eRD. Thus, considering the combined adiabatic power spectrum at the onset of standard RD, the power spectrum drops suddenly at $k=k_m$ leading to the first peak in the ISGWB spectra. Similar suppression is also expected for the isocurvature-induced adiabatic modes. The suppression for the isocurvature-induced adiabatic component is slower ($\propto 1/k^2$) than the inflationary adiabatic modes ($\propto 1/k^4$) \cite{Domenech:2020ssp}. Thus, the isocurvature-induced adiabatic component dominates for modes with $k> k_m$ and leads to the second peak in ISGWB at the frequency corresponding to the cutoff of the isocurvature power spectrum, $k_{\rm UV}$.

 The GW spectra are produced by both inflationary and the isocurvature induced adiabatic contributions for the scalar power spectrum described in section \ref{PPS}. As a result, each of the GW spectra has two peaks. Both are resonant contributions generated just after the evaporation of PBHs or at the start of standard RD. Inflationary perturbations contribute to the first peak, and the second peak is due to the isocurvature-induced adiabatic contribution from PBH distributions. In Fig.~\ref{sps-gw1}, we also limit the ratio between the conformal time at the beginning and at the end of PBH domination ($\tau_{rat}=\tau_{r}/\tau_{m}$). We stay in the regime where $\tau_{rat} \le 470 $ to avoid the breakdown of linear theory for scalar perturbations. For the validity of the linear theory, we require the duration of the early matter domination to be small ~\cite{Inomata:2019zqy, Kohri:2018awv, Assadullahi:2009nf}. This will constrain the initial mass fraction of PBHs $\beta_f$ we can probe. This effect, in particular, becomes significant for slightly heavier PBHs, as we shall see in later sections.

\section{Analytical approximation and signal to noise ratio}
\label{Anum}
In this section, we study the detectability of our signals in various detectors with the standard signal-to-noise ratio (SNR) approach. The spectrum, far away from the peaks, has a negligible impact on the SNR estimation. Thus we derive analytical formulas valid very close to the two resonant ISGWB peaks. Our calculation in this section closely follow previous works by Domenech et al~\cite{Domenech:2020ssp} and Inomata et all~\cite{Inomata:2019ivs}.

We are interested only in the resonant part of the induced gravitational wave, so following the calculation of the kernel term ($\overline{\mathcal{I}^2}$) from the appendix of \cite{Bhaumik:2020dor}, we take only the $\text{Ci}\left(\left(\sqrt{3} u+\sqrt{3} v-3\right) x_r/6\right)^2$ term :
\begin{align}
   \overline{\mathcal{I}^2} \approx \frac{x_r^8 \left(u^2+v^2-3\right)^4 \text{Ci}\left(\left(\sqrt{3} u+\sqrt{3} v-3\right) x_r/6\right)^2}{5971968 u^2 v^2} \, .
\end{align}
\begin{figure}[t!]
\begin{center}
\includegraphics[width=14cm]{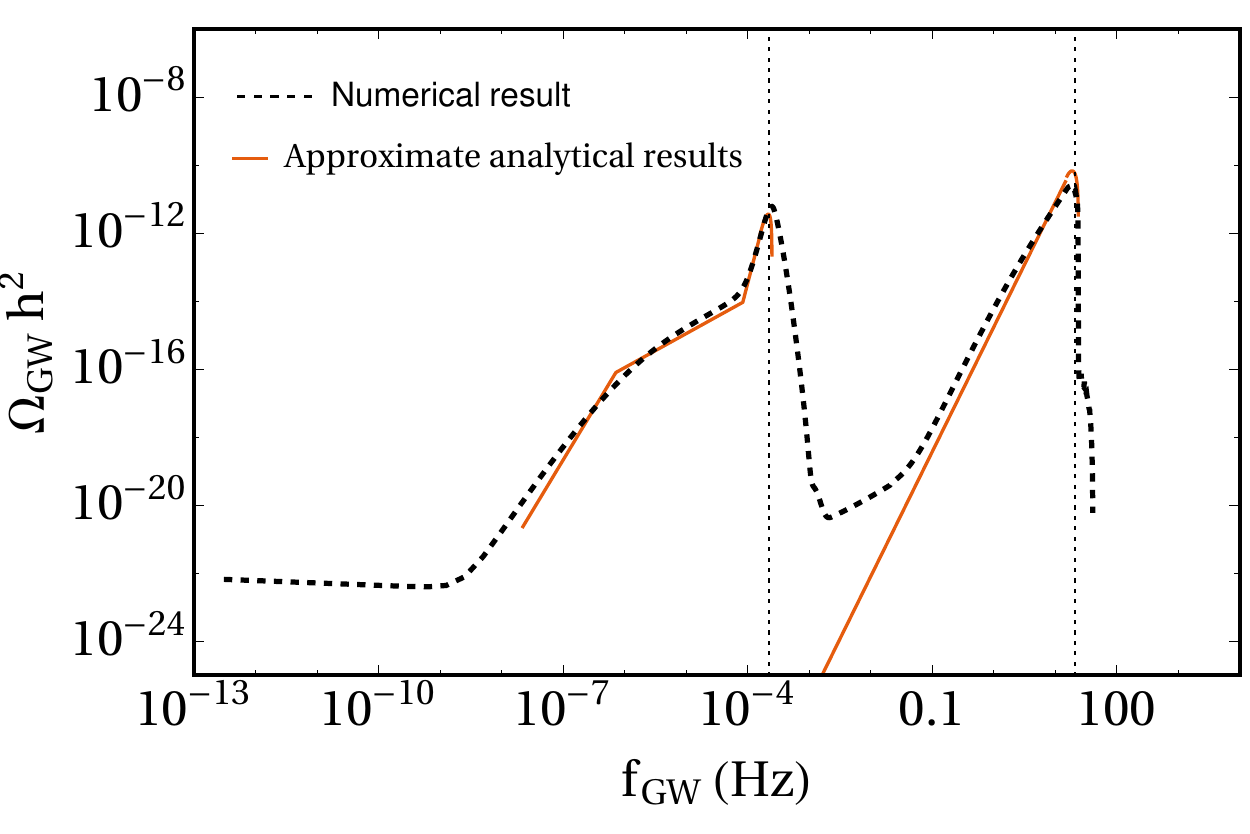}
\caption{\it Comparison of our analytical approximation with the numerical results for $M_{\rm PBH}=10^6$ g and $\beta_f=4.8 \times 10^{-8}$. Here two vertical dotted lines correspond to the frequencies associated with eRD-eMD era transition scale $k_m$ (in the left) and cutoff scale of isocurvature perturbations $k_{\rm UV}$ (in the right), respectively. As seen from the plot, these scales roughly correspond to two peaks originating from two different sources of scalar perturbations.}
\label{sps-gw2}
\end{center}
\end{figure}
This term plays the dominant role and leads to resonant amplification as the argument of Ci function approaches zero ( for $ u+v \approx \sqrt{3}$). With this simplified form of kernel, following the power spectrum calculation of section \ref{GWPBH} for the isocurvature contributed adiabatic perturbations we can express the tensor power spectrum as
\begin{align}
   \mathcal{P}_h= \int_{0}^{\infty}\dd v \int_{|1-v|}^{1+v} \dd u \frac{k_{m}^8 {x_r}^8 \text{Ci}(y)^2 \left(\left(-u^2+v^2+1\right)^2-4 v^2\right)^2}{14155776 ~2^{2/3} \sqrt[3]{3} \pi ^2\ ~u v~  k^2 k_{\rm UV}^6  \left(\frac{u k}{k_r}\right)^{2/3}
   \left(\frac{v k}{k_r}\right)^{2/3}}\, ,
\end{align}
 where we have taken $y\equiv (\sqrt{3} u+\sqrt{3} v-3) x_r/6$. We further redefine the variables using 
 \begin{align*}
     u\to {\sqrt{3} \left(\frac{s }{2 \sqrt{3}}+\frac{1}{2}+\frac{y}{x_r}\right)}, \hspace{1cm}
     v\to {\sqrt{3} \left(-\frac{s}{2
   \sqrt{3}}+\frac{1}{2}+\frac{y}{x_r}\right)} \, ,
 \end{align*}
 then integrating for the relevant limits we get the final form of the induced GW spectrum,
\begin{align}
\Omega_{GW}(\tau_0,k) & = c_g \,\Omega_{r,0} ~ \mathcal{J} \int_{-s_0}^{s_0} \frac{27 \sqrt[3]{3} \left(s^2-1\right)^2}{\left(9-3 s^2\right)^{5/3}} \dd s\\
 & =  c_g \,\Omega_{r,0} ~ \mathcal{J} \frac{2}{5} s_0 \left(\frac{3 \left(14-3 s_0^2\right)}{\left(1-\frac{s_0^2}{3}\right)^{2/3}}-37 \,
   _2F_1\left(\frac{1}{2},\frac{2}{3};\frac{3}{2};\frac{s_0^2}{3}\right)\right) \, ,
   \label{iso01}
\end{align}
where,
\begin{align*}
 \hspace{1cm}    \mathcal{J}=\frac{k^3 k_{m}^8 \left(\frac{k}{k_r}\right)^{2/3}}{1327104 \sqrt[3]{2} \sqrt{3} \pi k_r^5 k_{\rm UV}^6} \, .
\end{align*}
Here $_2F_1$ is the hypergeometric function. This result is obtained taking the integration of $y$ variable from $-\infty$ to $+ \infty$, which contributes a factor of $\pi$. For $k> 2/\sqrt{3} k_{\rm UV}$ we get a sharp cutoff in the GW power spectrum, but for $k< 2/\sqrt{3} k_{\rm UV} $, the limit of $s$ integral is a function of k \cite{Domenech:2020ssp},
\begin{align}
s_0=\left\{
\begin{aligned}
&1\qquad &\tfrac{k}{k_{\rm UV}}\leq\tfrac{2}{1+\sqrt{3}}\\
&2\tfrac{k_{\rm UV}}{k}-\sqrt{3}\qquad & \tfrac{2}{1+\sqrt{3}}\leq\tfrac{k}{k_{\rm UV}}\leq\tfrac{2}{\sqrt{3}}\\
\end{aligned}
\right.\, .
\end{align}
Here the second limit is valid very close to the spectrum's peak, and the first limit describes the spectrum at lower frequencies. The $s$ integral for the first limit gives a value close to $1.16$, while for the second limit it gives $ \approx 0.5$ at $k=k_{\rm UV}$.
\begin{figure}[h!]
\begin{center}
\includegraphics[width=14cm]{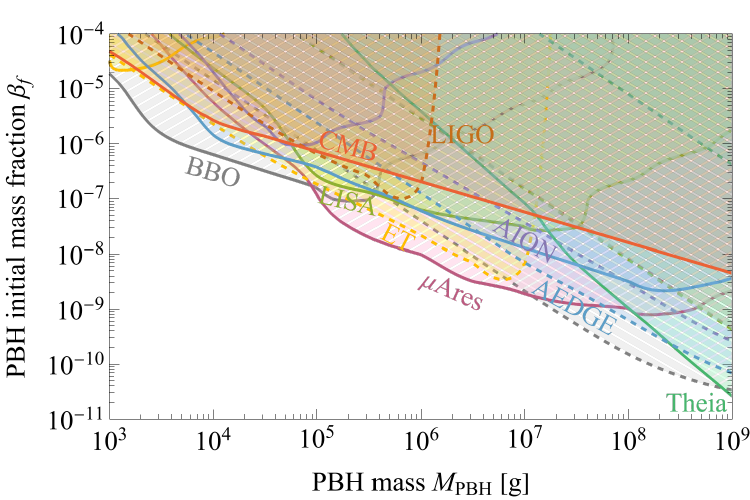}
\caption{\it Contours indicating the spectra within the detection range of future experiments through the standard criterion SNR $\geq 10$. Detection of each of our two signals is indicated separately for each experiment: detection of the low frequency inflationary adiabatic contribution is indicated by solid contours and $/$ dashed filling, while detection of the high-frequency isocurvature induced adiabatic contribution is indicated by dashed contours and $\backslash$ dashed filling. In all of the parameter spaces where both contributions are visible in a combination of experiments, we could easily distinguish our unique spectrum and the model at hand. Finally, the solid red contour shows the part of the parameter space excluded by the overproduction of GWs spoiling the CMB~\cite{Henrot-Versille:2014jua, Smith:2006nka}.}
\label{fig:sps-gw2}
\end{center}
\end{figure}
Similarly it is also possible to obtain an analytical understanding of induced $\Omega_{GW}$ near the first peak or inflationary scalar perturbations induced peak for standard power law inflationary power spectrum, defined in \eqref{power-law}. The cutoff scale for inflationary power spectrum is not relevant here, because any scale which enters the horizon during eRD gets sufficiently suppressed leaving an effective cutoff scale at $k=k_{m}$. We take only the resonant contribution very close to the peak~\cite{Inomata:2019ivs} ,
\begin{align}
\frac{\Omega_{GW}(\tau_0,k) }{A_{\text{s}}^2 c_g \,\Omega_{r,0}} \simeq & 
\begin{cases}
3 \times 10^{-7} x_r^3 x_{\text{max}}^5  &   150 x_{\text{max}}^{-5/3} \lesssim x_r \ll 1 \\
6.6 \times 10^{-7} x_r x_{\text{max}}^5 & 1 \ll x_r \lesssim x_{\text{max}}^{5/6} \\
3 \times 10^{-7}   x_r^7 &     x_{\text{max}}^{5/6} \lesssim x_r   \lesssim \tfrac{2}{1+\sqrt{3}}x_{\text{max}}   \\
\mathcal{C}(k)  &  \tfrac{2}{1+\sqrt{3}}\leq\tfrac{x_r}{x_{\text{max}}}\leq\tfrac{2}{\sqrt{3}}\\
 \end{cases} \, ,
 \label{ad01}
\end{align}
where 
\begin{eqnarray}
 \mathcal{C}(k) =0.00638 \times 2^{-2 n_s-13} ~3^{n_s} ~x_r^7 ~s_0
\left(\frac{x_r}{x_{max}}\right)^{2 n_s-2} \times \hspace*{5cm}\nonumber \\
 \left(-s_0^2 \, _2F_1(\frac{3}{2},-n_s;\frac{5}{2};\frac{s_0^2}{3})   +4 \, _2F_1(\frac{1}{2},1-n_s;\frac{3}{2};\frac{s_0^2}{3})-3 \,  _2F_1(\frac{1}{2},-n_s;\frac{3}{2};\frac{s_0^2}{3}) \right) \, .
\end{eqnarray}
Here also the limit of the $s$ integral is similar to isocurvature contributed peak. $s_0 = 2\frac{k_{m}}{k}-\sqrt{3}$ and $x_r=k/k_r$, $x_{max}=k_{m}/k_r$ are evaluated at $\tau=\tau_r=1/k_r$. 

We use these results to determine the detection possibilities by robustly calculating SNR for different future GW detectors for different mass and initial abundance of PBHs. In each case, we use the noise curve of the given experiment and calculate
\be
{\rm SNR} \equiv \sqrt{\mathcal{T}\int {\rm d}f\, \left[\frac{\Omega_{\rm GW}(f)}{\Omega_{\rm noise}(f)}\right]^2} \,,
\ee
\begin{figure}[h!]
\begin{center}
\includegraphics[scale=0.7]{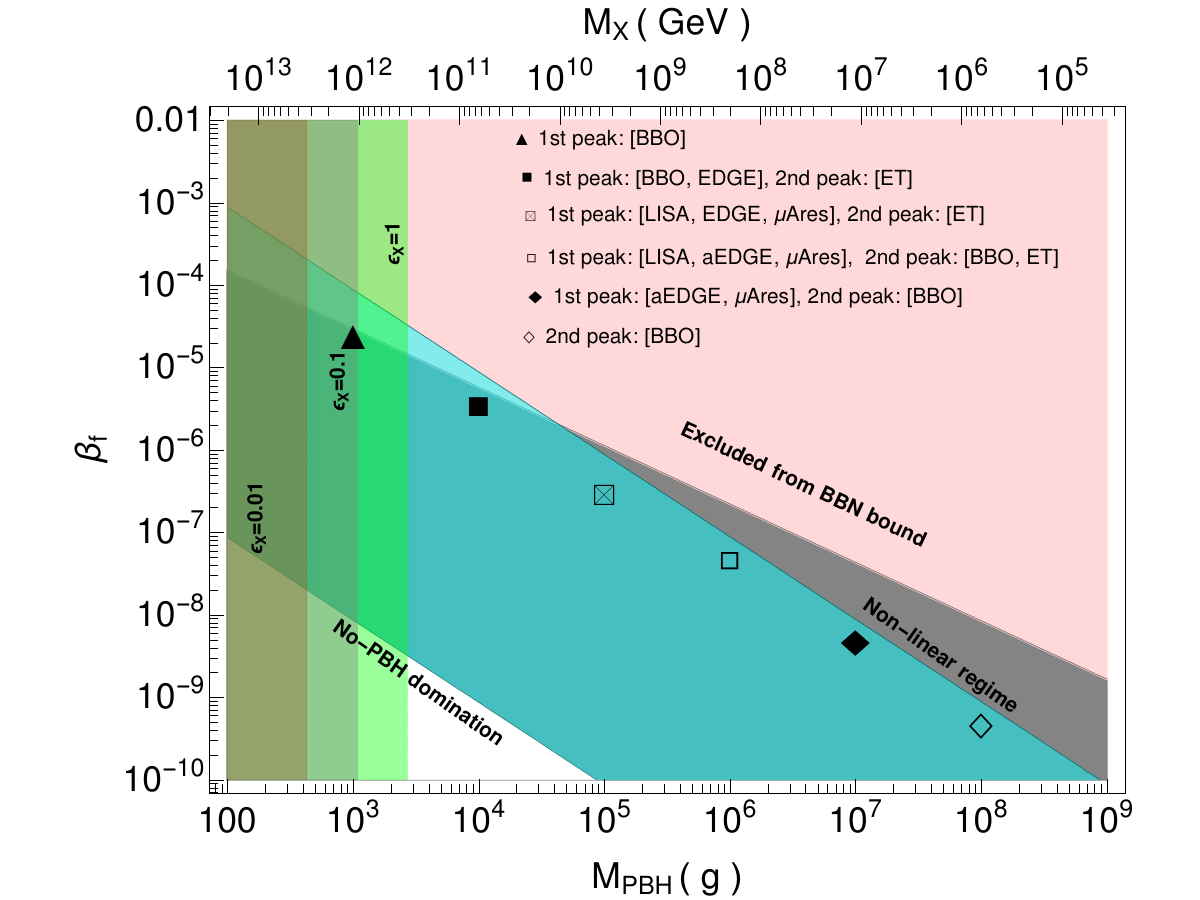}
\caption{\ Here, we plot constraints on mass and initial abundance of PBHs from different aspects. The benchmark points in the plot correspond to the peaks shown in Fig. \ref{sps-gw1}, the light-red region is excluded from BBN bound, and the cyan region corresponds to the finite duration of PBH domination. The lower bound on the cyan patch corresponds to the cases where PBH evaporates before domination, and the upper bound comes from the non-linearity bound ($\tau_{rat} \le 470$). In the left side, the light-green, green, and grey vertical contours represent the parameter space allowed for GUT baryogenesis models satisfying the proton decay bound for Higgs triplet particles ($M_\mathcal{T} \ge 10^{11}$ GeV) for efficiency parameter $\epsilon_X= 1, 0.1, \text{and } 0.01 $ respectively. The upper bar indicates the associated mass range of heavy right-handed neutrinos for baryogenesis via the leptogenesis mechanism, while we stick to efficiency parameter $\epsilon_X=1.0$. }
\label{con2}
\end{center}
\end{figure}
assuming operation time $\mathcal{T}=4$~yr for each experiment. We display the results for the entire parameter space of interest in Fig.~\ref{fig:sps-gw2}. We also include an upper bound coming from the overproduction of GWs spoiling the CMB~\cite{Henrot-Versille:2014jua, Smith:2006nka} and separate the SNR to that coming from detection of the low and high-frequency peak indicated by $/$ and $\backslash$ slanted dashed filling. Thus we also specify the areas in which both peaks are visible, and our smoking gun signal would be distinguishable through a combination of measurements from various GW experiments. 

\section{Combined constraints from baryogenesis and ISGWBs}
\label{combined}
There has been no confirmed detection of stochastic GW background so far, and as a result, the only way to constrain the parameter space is through the BBN bound on the GW background. The GW energy density during BBN must be smaller than twenty percent of the photon energy density : $\Omega_{GW}(\tau_{BBN}) \le 0.05 $  or  $\Omega_{GW}(\tau_{0}) \le 0.05 ~ c_g ~\Omega_{r,0} $ ~\cite{Caprini:2018mtu}. 
The non-linearity bound ($\tau_{rat} \le 470$) constrains the first peak or the peak from inflationary perturbation to an upper limit, $\Omega_{GW}(\tau_{0}) \lesssim 10^{-10}$, which is far lower than BBN bound. A similar thing also happens for isocurvature-induced adiabatic contribution coming from higher mass PBHs. But as we go towards lower mass PBHs, even for a small value of $\tau_{rat}$ ($ < 470$), the isocurvature contribution can be quite large, and it can potentially constrain the initial abundance of primordial black holes in terms of initial mass fraction $\beta_f$. This trend is visible from Fig.~\ref{sps-gw1} and the contour plot in Fig.~\ref{con2}.

As the amplitude of the isocurvature induced second peak of the GW spectrum is the only relevant quantity for BBN bound, we use the analytical estimation of this second peak amplitude at BBN epoch, $\Omega_{GW}(\tau_{BBN})$
to constrain the parameter space, namely the mass and initial abundance of PBHs in Fig.~\ref{con2}. After the source term stops to contribute $\Omega_{\rm GW}$ stays nearly constant throughout the RD, and we can assume $\Omega_{GW}(\tau_{BBN}) \approx \Omega_{GW}(\tau)$, defined in equation \eqref{RD-Omega}. We also get a lower bound on $\beta_f$ for $\tau_{rat} \ge 1$, because PBHs shall evaporate before they can dominate for very small values of $\beta_f$ (the white region at the down-left corner in Fig.~\ref{con2}). We are not considering the parameter space where PBHs evaporate before they can dominate the universe, as this parameter space shall not lead to any resonant amplification in the ISGWB.

In the context of GUT baryogenesis, due to the proton lifetime bound on the mass of the Higgs triplet particles ( $M_{\mathcal{T}} \ge 3 \times 10^{11} GeV$ ) only very low mass PBHs can contribute to consistent baryogenesis process ($Y_B=8.8 \times 10^{-11}$ ). We are also plotting this bound on the mass of PBHs for different values of $\epsilon_X$ in Fig.~\ref{con2} with vertical contours of color grey, green, and light green, respectively. Here $\epsilon_X$ is the efficiency parameter for asymmetric baryon decay of the intermediate Higgs triplet particle, defined in equation~\eqref{epsilon}. 

The proton decay bound is absent for baryogenesis via leptogenesis process, and the mass range of the right-handed neutrinos is essentially unconstrained. The only constraint on this scenario is the sphaleron washout, which places an upper limit on the PBH mass~\cite{Hooper:2020otu}. Fig.~\ref{con2} shows the mass range of right-handed neutrinos on the upper axis, with the assumption $\epsilon_X = 1$. This mass range would be modified significantly for $\epsilon_X < 1 $. From equation~\eqref{MX} we can see that the relevant mass of the decaying particles increases as we decrease the initial mass of PBHs.

One interesting point to note here is that the connection of baryogenesis to the ISGWB is through the ultra-low mass PBH distribution. The baryogenesis from the PBH evaporation scenario shall always lead to this detectable ISGWB signature. But, the ISGWB signature is not exclusive to the baryogenesis scenario. We can expect the amplification in the ISGWB spectrum whenever ultra-low mass PBHs with a sharp monochromatic mass distribution dominate the universe for a short duration before standard RD.


\section{Conclusions and discussions  }
\label{discuss}

In this paper, we investigated the ISGWB signature of non-thermal baryogenesis induced by PBH evaporation as a pathway to address the matter-antimatter asymmetry of the universe. Due to its non-thermal nature, the dark sector is never thermalized in the early universe. Since there is no interaction between the dark and visible sectors, it is usually difficult to test such baryogenesis mechanisms in laboratory experiments or astrophysical searches. Here we propose to test such baryogenesis mechanisms via induced gravitational waves background, which originate from the second-order tensor perturbations induced by first-order scalar perturbations. In particular, we emphasize the resonant part of the GW signals generated just after the sudden transition from PBH-dominated eMD to standard RD. We analyze two contributions of the GW spectrum: the isocurvature induced adiabatic perturbation assuming the Poisson distribution of PBHs, and the inflationary adiabatic curvature perturbation assuming the power-law power spectrum. This results in a unique doubly-peaked power spectrum and predicts detectable signals at various current and upcoming GW observatories. We investigate how these signals may shed light on the baryogenesis mechanism from Hawking evaporation of PBHs. We summarize the main findings of our analysis below:
\begin{itemize}

    \item For PBH mass range $10^2$ g to $10^8$ g, we estimated the resonant ISGWB, contributed from the isocurvature induced adiabatic perturbations associated with PBH distribution and from the inflationary adiabatic scalar perturbations, as shown in Fig. \ref{sps-gw1}.
    
    \item The unique shape of the GW spectrum we find would serve as a smoking-gun signal for our scenario, making possible \textit{one-to-one} correspondence between the GW spectrum and baryogenesis. This allows us to distinguish it from other non-thermal baryon asymmetries production mechanisms. Interestingly, isocurvature mode can also be generated for the baryogenesis from Q-ball in the Affleck-dine\cite{White:2021hwi}  scenario. Thus, the only difference in the ISGWB spectrum shall come from the different lifetime and decay rates of Qballs compared to PBHs. If varying the Qball mass and other parameters offer degeneracy with corresponding PBH lifetime and evaporation rate, the ISGWB spectrum shall also have a similar shape. But in these two cases, the details of baryogenesis parameters shall be very different even for an identical ISGWB spectrum.
    
 \item We provide analytical expressions (equation \eqref{iso01} and \eqref{ad01}) for the ISGWB spectrum, valid near the peaks of the signals. We find it in good agreement with our numerical estimation (See Fig. \ref{sps-gw2} for comparison). 
    
\item We examine the detection possibilities of our doubly-peaked GW signals for different mass ranges and initial abundance of PBHs, (with $\text{SNR} \geq 10$), as shown in Fig.~\ref{fig:sps-gw2}.

\item  We also estimate the limiting initial abundance of PBHs imposing the BBN bound. This becomes more relevant for black hole mass range $10^2-10^4$ gram. For heavier PBHs, the initial abundance is stringently restricted for the validity of our linear perturbation formulation ($\tau_{rat} \leq 470$). We discuss these bounds in section \ref{combined} with  Fig.~\ref{con2}.
    
\item For the observed value of $Y_B$, the mass of the heavy decaying particle $M_X$, becomes the function of $\epsilon_X$ and $M_{\rm PBH}$. For the GUT theories involving Higgs triplets, its mass is restricted from proton decay bound $M_X \ge 3 \times 10^{11} GeV$. This allows only low mass PBHs to generate a feasible amount of baryogenesis. Thus any detection of our characteristic SGWB signal associated with $M_{\rm PBH} > 10^4 ~{\rm g}$ will favor the second scenario where right-handed neutrinos decay leading to baryogenesis through leptogenesis. We show the corresponding mass range of PBHs for these two cases in Fig.~\ref{con2}. 

\item For a broader mass distribution of PBHs, the transition from PBH domination to radiation domination can not be assumed instantaneous as different mass PBHs shall evaporate at slightly different times. A prolonged transition from eMD to RD suppresses both inflationary and isocurvature-induced adiabatic modes. This suppression has an exponential dependence on the wavenumber \cite{Inomata:2020lmk} and the isocurvature-induced peak comes at a higher wavenumber. Thus it is expected to have more significant suppression than the inflationary adiabatic part.

\end{itemize}
We envisage our results to provide a pathway between the detection of GW signatures from PBH on the one hand and the generation of baryon asymmetry on the other hand. This becomes particularly desirable and very important in high-scale baryogenesis and leptogenesis, as such scenarios are otherwise not within reach of laboratory physics experiments. Our analysis paves the way for BSM model building to realize viable scenarios for such high-scale baryogenesis mechanisms in reach of upcoming experiments, including GUT-scale, SUSY, and heavy right-handed neutrino-based baryogenesis mechanisms. We are limiting ourselves to PBHs with mass greater than $10^2$g. With the advent of ultra-high frequency, gravitational wave detection techniques \cite{Franciolini:2022htd, Aggarwal:2020umq} even lower mass PBHs will be relevant to explore the ISGWB above kHz frequency range. We plan to address additional interesting possibilities, such as going beyond the initial Poisson distribution of PBHs and incorporating the effects of PBH clustering or non-Gaussian scalar perturbations in future publications.


\section*{Acknowledgment}
NB thanks Rajeev Kumar Jain for many useful discussions at various stages of this manuscript. Authors also thank Guillem Domènech, Ranjan Laha, P. Jishnu Sai, Yashi Tiwari and Arnab Paul for very helpful discussion and suggestions. NB acknowledges financial support from the Indian Institute of Science(IISc), Bangalore, India, through full-time research fellowship. This work was supported by the Polish National Science Center grant 2018/31/D/ST2/02048. ML was also supported by the Polish National Agency for Academic Exchange within Polish Returns Programme under agreement PPN/PPO/2020/1/00013/U/00001. 


\bibliographystyle{JHEP}
\bibliography{bib2}
\end{document}